\documentstyle[aps,preprint]{revtex}
\tightenlines
\begin{document}

\title{On decay of bubble of disoriented chiral condensate}

\author{V. A. GANI \thanks{E-mail: gani@vitep5.itep.ru}}

\address{
Moscow State Engineering Physics Institute (Technical University),\\
Kashirskoe shosse, 31, Moscow, 115409, Russia\\
and\\
Institute of Theoretical and Experimental Physics, Russia\\
}

\author{A. E. KUDRYAVTSEV \thanks{E-mail: kudryavtsev@vitep5.itep.ru},
T. I. BELOVA and B. L. DRUZHININ}

\address{{\it Institute of Theoretical and Experimental Physics,}\\
{\it B.Cheremushkinskaya, 25, Moscow, 117259, Russia}\\
}

\maketitle

\begin{abstract}
We discuss the space-time structure for the process of decay of a
bubble of hypothetic phase~-- disoriented chiral condensate (DCC). The
evolution of the initial classical field configuration corresponding to the
bubble of DCC is studied both numerically and analytically. The process of
decay of this initial configuration depends crucially on selfinteraction of
the pionic fields. It is shown that in some cases this selfinteraction leads
to the formation of sort of breather solution, formed from pionic fields
situated in the center of the initial bubble of DCC. This breather looks like
a long-lived source of pionic fields.
\end{abstract}

\section{Introduction}

     Disoriented chiral condensate (DCC) is nothing more than a piece of
vacuum, chirally rotated from its usual orientation in internal symmetry
space. The history of the subject is rather old
\cite{Einkova,Andreev,Karmanov,Lam}.
Formation of a large domain with DCC has been proposed by
Anselm~\cite{Anselm_PRL}, by Blaizot and Krzywicki~\cite{Blaizot} and by
Bjorken, Kowalski and Taylor~\cite{Bjorken}. The more detailed lists of
references devoted to this subject one may find in the more recent
publications, see, e.g. Refs.~\cite{Amado,Anselm_ZP}.

     Usually people discuss the radiation of pionic fields from initially
formed bubble of DCC. The most important feature of this radiation is
semiclassical coherent nature of the initial pionic configuration. As usual,
the four fields $\sigma$ and $\vec{\pi}$ are constrained by the relation
$$
\sigma^2+\vec{\pi}^2=f^2_{\pi} \ . \eqno(1)
$$
Usually one supposes that in ordinary vacuum $<\sigma>=f_{\pi}$. The
DCC-hypothesis means that inside the DCC bubble the chiral vacuum does not
point in the $\sigma$ direction but rather in one deflected toward one of
the $\pi$ directions.

     If the event-by-event deviation of the chiral orientation from the
$\sigma$ direction is random, it follows that the distribution of neutral
fraction
$$
f=\frac{N_{\pi^0}}{N_{\pi^+}+N_{\pi^-}+N_{\pi^0}}=\frac{N_{\pi^0}}{N}
\eqno(2)
$$
is inverse square root, as $N$ becomes large:
$$
\frac{dP}{df}=\frac{1}{2\sqrt{f}} \ . \eqno(3)
$$
At the same time the "standard" non-coherent mechanisms (like we get at high
energies) predict the probability
$$
\frac{dP}{df}=\delta(f-1/3) \ . \eqno(3a)
$$
The distribution (3) is the most bright and evident signal of DCC-phase.

     It would be interesting to find out what are other consequences in terms
of observables for decay of the DCC bubble. In other words, what is the
space-time picture of the process of a DCC bubble decay? To answer this
question, let us consider the simplified model with only a pair of fields
$\sigma$ and $\pi$ under constraint (1). In this case the choice
$<\sigma>=f_{\pi}$ means "ordinary" vacuum state. It is more suitable to
introduce new angular variable $\phi$: $\pi=f_{\pi}cos\phi$,
$\sigma=f_{\pi}sin\phi$. In terms of this new field $\phi\in[0,2\pi]$ vacuum
is degenerate. So we have $U(1)$ symmetry in the vacuum sector.

     It is evident, that if we are limiting ourselves by noninteractive
fields only, the evolution of any initial field configuration in terms of
$\phi$ field is to satisfy the equation
$$
\phi_{tt}-\Delta\phi=0 \ . \eqno(4)
$$
At the same time one may introduce some violation of the chiral symmetry, by
adding mass of "pion" in this equation, namely
$$
\phi_{tt}-\Delta\phi+m^2\phi=0 \ . \eqno(5)
$$
This assumption is equivalent to the condition that the theory has only one
real vacuum, $<\sigma>=f_{\pi}$, $<\pi>=0$ and any other position on the
circle $\sigma^2+\pi^2=f_{\pi}^2$ is not the real vacuum state.
Nevertheless, the bubbles with nonzero expectation value of $\phi$ could be
initially formed, and we may discuss properties of decay of such states.
One may consider also a more complicated case of the structure of vacuum,
considering the situation with more than one vacuum state. This situation may
be approximated by the following equation of motion:
$$
\phi_{tt}-\Delta\phi+\frac{1}{2}m^2\sin2\phi=0 \ . \eqno(6)
$$
There are two vacuum states in this theory with $\phi=0$ and $\phi=\pi$
($0\le\phi\le2\pi$).
This form of potential in terms of the fields $\sigma$ and $\pi$ resembles a
Mexican folk hat though arranged in the Texas style with back and front brims
curved down.

     As we shall demonstrate, the decay of the DCC bubble looks quite
different in these three cases (4), (5) and (6). That is why it is possible
to get information about real interaction in the system studying the decay of
DCC bubbles. Evidently, cases (4)-(6) give us only some examples of
selfinteraction and the physical picture may be more complicated.
Nevertheless, limiting ourselves by these examples, we get some general
features of the decay of DCC.

\section{One-dimensional case}

     Let us consider first for pedagogical reason the evolution of the
initially formed bubble of DCC in (1+1)-dimensions.

     {\bf a) Massless case}. The solution of the equation of motion (4) is
trivial in this case
$$
\phi(x,t)=\frac{1}{2}[\phi_0(x-t)+\phi_0(x+t)] \ , \eqno(7)
$$
where $\phi_0(x)$ is the initial field configuration. Taking the initial
field configuration in the form
$$
\phi(x,0)\equiv\phi_0(x)=\frac{\phi_0}{1+(x/a)^{2\alpha}} \ ,
\ \phi_t(x,0)=0 \ , \eqno(8)
$$
$$
\alpha\gg1\ {\rm is\ positive\ integer},
$$
we just immediately get from Eq. (7) that a detector being
placed at the distance $X\gg a$ from the position of the domain of DCC, will
fix the flux outgoing from the region of DCC after the time interval
$t\sim X/c$. The signal lasts the period $\tau\sim2a/c$ and consists of two
separate peaks corresponding to the forward and backward walls of the
initial bubble passing across the detector. Concentration of energy in the
boundaries reflects only the fact, that vacuum is exactly degenerate, so all
the excess of energy of the DCC bubble is concentrated on its surface due to
the gradient terms in the expression for energy density of the system.

     {\bf b) Massive case}. The situation is getting only slightly more
complicated in the massive case (5). In this case the solution of the problem
can also be obtained analytically as Eq. (5) is linear. Time evolution of the
initial configuration $\phi_0(x)$ (8) depends on the value of the
dimensionless parameter $\xi=ma$.

     The case $\xi\ll1$ looks very similar to the massless case with the only
exception. The interior region of the bubble carries also some part of energy
(volume energy). This is the consequence of the chiral symmetry violation in
the vacuum sector. The solution in the interior region oscillates in time.
Also it is worth mentioning, that practically all the flux of energy is
transferred with near-to-light velocity. There present also some more slowly
moving tail, but its energy is negligibly small under condition $\xi\ll1$.
The situation changes with increasing $\xi$. Part of energy, carried with
near-to-light velocities, is decreasing. At the same time the process of
emitting of slowly moving particles dominates. The flux of energy through
the point $X=20$ as a function of time is represented in Fig. 1 for several
values of $\xi$. Obviously, that $\xi=0$ corresponds to the massless case (4).

     {\bf c) sine-Gordon case}.
     The situation is getting more complicated in the case of dynamics with
degenerate vacua, described by Eq. (6). In this case the process of bubble
evolution depends on the initial mean value of $\phi$ in the interior region
of the bubble.

i) Consider first the case of small $\phi_0$: $|\phi_0|\ll1$. In this case
the decay of the bubble is similar to one in massive case b). The only
difference is in formation in the center of the bubble a sort of breather
solution of small amplitude. This breather solution is a well-known
breather of the sine-Gordon equation~\cite{Perring}. It is stable and it
keeps part of the initial energy of the bubble in the center. So not all
the initial energy of the DCC is emitted.

ii) In the case $|\phi_0-\pi|\ll1$ the field oscillations take place already
not around $\phi=0$, as for $|\phi_0|\ll1$, but around $\phi=\pi$ (second
local minimum in the system). If we take initial field configuration in the
form of (8) with $\phi_0=\pi$, $\xi=ma\gg1$, we get that for large times the
field configuration will look like a far situated kink-antikink pair
$$
\phi_{K\bar{K}}\approx2\arctan\exp[m(x+a)]-2\arctan\exp[m(x-a)]
$$
with some small oscillations localized in the center near $x=0$. These
oscillations are due to the initial condition (8) for field configuration and
they leave the region of interaction relatively fast. If $|\phi_0-\pi|\ne0$,
we observed the formation of a breather-like solution of small amplitude in
the central region. We would like to emphasize, that in the selfinteraction
case considered above part of energy, concentrated initially in the DCC
bubble, is not emitted at all but forms special stable localized solution
called breather. Conservation of part of energy of the DCC inside the bubble
is indeed the direct consequence of integrability of Eq. (6) in
one-dimensional case.

\section{Three-dimensional case}

     Solution of Eq. (4) in spherically symmetric three-dimensional case is
also trivial. Really, making substitution $\phi(r)=w(r)/r$, we get
$$
w_{tt}-w_{rr}=0 \ , \eqno(9)
$$
so the solution is known analytically, see Eq. (7). The situation looks
analogously in the massive case (5), and we get for $w(r,t)$ in this case:
$$
w_{tt}-w_{rr}+m^2w=0 \ . \eqno(10)
$$

     The situation with the sine-Gordon equation looks more complicated in
three-dimensional case, namely we get for $w(r,t)$ the following equation:
$$
w_{tt}-w_{rr}+\frac{1}{2}m^2r\sin(2w/r)=0 \ . \eqno(11)
$$
Later on we shall consider the bubbles according to Eqs. (9)-(11) with
initial conditions, chosen in the form
$$
\phi(r,0)=\frac{\phi_0}{1+(r/a)^{2\alpha}} \ , \ \phi_t(r,0)=0 \ , \eqno(12)
$$
where $\alpha$ is positive integer, as in (8). The behavior of the energy
flux in cases (9)-(11) is shown in Fig. 2. The general picture looks similar
to that of one-dimensional case. Nevertheless, some comments are needed
when comparing results for the sine-Gordon equation in one-dimensional and
three-dimensional cases. In one-dimensional case the breather solution is
stable. In three-dimensional case with $|\phi_0|\ll1$ numerically we also
observed formation of sort of breather in the center of the initial bubble.
This three-dimensional breather was already found long ago~\cite{Bogol_JETP}.
It is known,
that this solution is quasi-stable and it decays slowly, emitting
radial waves. That is why the process of decay of the DCC with dynamics
corresponding to Eq. (11) has two stages. During the first stage the main
part of energy is emitted. This first stage is finished by the formation of
breather in the center of the initial bubble. Afterwards the emission of
waves becomes a slow process. We solved Eq. (11) also with the following
initial condition:
$$
\phi(r,0)=2\arctan\exp[-m(r-a)] \ , \ \phi_t(r,0)=0 \ . \eqno(13)
$$
Initial condition (13) corresponds to the minimum of energy of the field
which links two vacua of the theory. Initial condition (12) does not
correspond to this minimization procedure. It has additional excess of
energy, especially if $\alpha\gg1$. This excess of energy may be considered
as excitation over soliton (13). In the sine-Gordon case this excitation
belongs to the continuum and may be emitted from the soliton region. Thus,
the evolution of the bubble is similar to the case of initial condition of
the type of (13) with the only difference. The initially formed bubble with
$\phi=\pi$ vacuum inside ($\phi_0=\pi$ in (12)) is not only collapsing, but
emitting radial waves from its boundary too.

     The time dependence of the field $\phi(r,t)$ at the origin and the flux
of energy from this solution under initial condition (13) are represented in
Fig. 3. The time dependence
of $\phi(0,t)$ may be conventionally separated into two periods. The first
one $0<t\le T_0$ is characterized by oscillations of large amplitude and
large damping. The second period $t\ge T_0$ ($T_0\sim200$ in our units)
may be characterized by oscillations with much smaller amplitudes (breather).
The damping is also small at this stage of evolution.

     The ratio of energies being emitted during the first fast
(gross-structure) and the second slow (breather) evolution stages of the
decay depends on the initial configuration of the bubble. It is worth
mentioning that the evolution of a spontaneously formed spherical bubble
was studied long ago in papers~\cite{Okun} for $\lambda\phi^4$-field theory.
The main conclusion of those papers was that the bubble of "wrong" vacuum
with the initial radius $R_0\gg1$ starts to collapse. The collapse lasted
the time $T\simeq1.3R_0$, which means that the walls of the bubble are moving
toward the center at nearly the speed of light.

     The pulsations of large amplitude (gross-structure) were found for
$\lambda\phi^4$-theory in a narrow vicinity of the initial radius $R_0=3.875$
\cite{Belova_JETP,Belova_YF}, see also review~\cite{Review}. In contrast to
the case of $\lambda\phi^4$-theory for the sine-Gordon model we observed the
gross-structure shown in Fig. 3 in the wide range of the initial size of the
bubble:
$5\le a\le50$ for $m=1$. We have not performed calculations for larger $a$,
but there seems to be the same gross-structure for all $a\gg1$. For smaller
$a\le5$ we observed the formation of a breather solution just after shrinking
of the initial configuration.

     So if we consider the picture of decay of the DCC bubble of small
amplitude, with $|\phi_0|\ll1$, it expands and nothing drastical happens.
But in the case of decay of bubbles with large amplitudes $|\phi_0-\pi|\ll1$,
the picture of decay looks quite different and it is similar to the
$\lambda\phi^4$-theory case, studied long ago.

\section{Conclusion}

     We studied the space-time structure for the decay of DCC bubble. The
main point is that characteristics of this decay depend crucially on
selfinteraction in the system of pionic fields. Namely, presence of
selfinteraction (nonlinear effects) leads to the formation of sort of
breather solution in the center of the initially formed bubble of DCC. This
breather solution decays slowly and keeps part of energy in the central
region. That is why the decay of the bubble is not instant ($\tau\sim a/c$),
but rather long process. The detailed picture of this decay depends on
details of selfinteraction in the system as well as on the initial conditions
(amplitude and size of the bubble.) That is why studying experimentally
the space-time structure of the signal followed the decay of the DCC
bubble, one may get limitations on the form of selfinteraction potential in
terms of chiral fields.

     As the decay of the DCC bubble may be experimentally observed by fixing
the flux of outgoing pions, it is worth to stress once again, that this flux
is to follow the time dependence drawn in Fig. 3b. So we may find not only
the prompt signal from decaying DCC, but also some delay pions emitted via
the breather decay.

     Notice, that in spite of the fact that we studied the most trivial case
of the two-component $\sigma$-theory, our result is indeed more general.
Really, one may consider theory with three pions and one sigma. In this case
vacuum sector is $S^3$-sphere. But the evolution of the system will take
place along the "large circle" of this $S^3$-sphere. That is why the time
evolution of the initial field configuration looks like in $S^1$-case,
studied in this our paper.

     In this paper we didn't study the influence of selfinteraction on charge
distribution. We think, that this distribution is to be independent on this
interaction and is given by Eq. (3).

     It is also worth mentioning, that the space-time structure of signal from
decay of the DCC bubble was the subject of discussion in
papers~\cite{Abado,Amelino}. The authors of paper~\cite{Abado} consider the
evolution picture for DCC phase in terms of linear sigma model.
In paper~\cite{Amelino} decay of the DCC bubble with both linear and
nonlinear $\sigma$-models was studied. In contrary to
papers~\cite{Abado,Amelino} in this our paper we concentrated on the
formation of a long-lived breather-like solution in the center of the bubble.
The existence of this solution is a consequence of selfinteraction of the
fields in vacuum sector.

\section*{Acknowledgments}

     The authors are thankful to Dr. N.~A.~Voronov and Dr. V.~G.~Ksenzov for
helpful discussions.

     This work was partly supported by the Russian Foundation for Basic
Research under grant No.~95-02-04681.

\begin{center}
\bf
FIGURE CAPTIONS
\end{center}

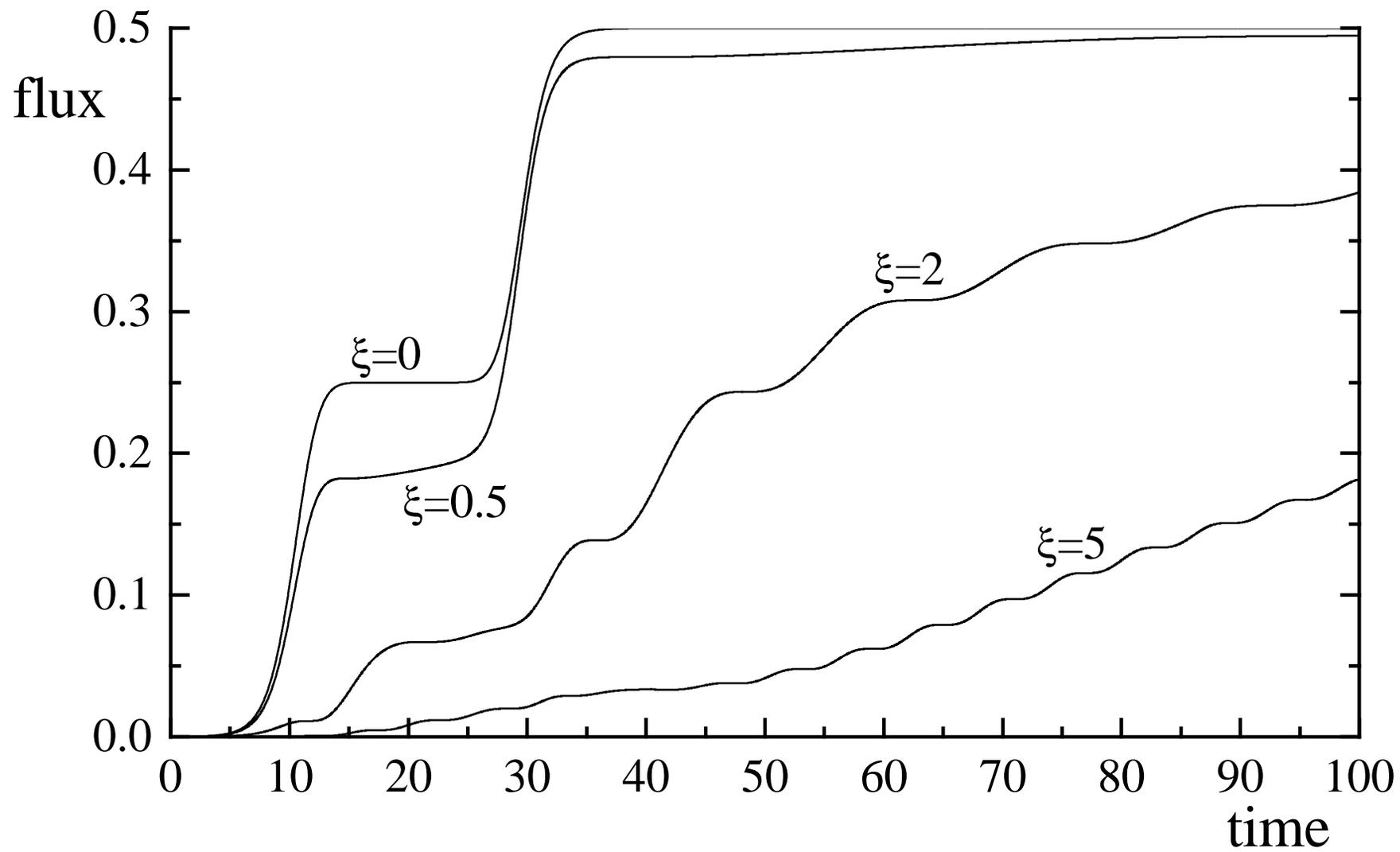
{\bf Fig. 1.}
Flux of energy (in units of total energy) from DCC bubble in one-dimensional
case for different $\xi=ma$ as a function of time ($\alpha=3$).

\bigskip

{\bf Fig. 2.}
Flux of energy as a function of time for three-dimensional case:\\
a) Eq. (10) with initial conditions (12), $\alpha=5$, $a=1$, flux through the
sphere $r=4$;\\
b) Eq. (11) with initial conditions (12), $\alpha=25$, $a=10$, $\phi_0=0.1$,
flux through the sphere $r=20$;\\
c) Eq. (11) with initial conditions (12), $\alpha=25$, $a=10$, $\phi_0=\pi$,
flux through the sphere $r=20$.

\bigskip

{\bf Fig. 3.}
Three-dimensional sine-Gordon case (11) with initial conditions (13),
$a=20$, $m=1$:\\
a) time dependence of the field $\phi(r,t)$ at the origin $r=0$;\\
b) flux of energy as a function of time in this case.

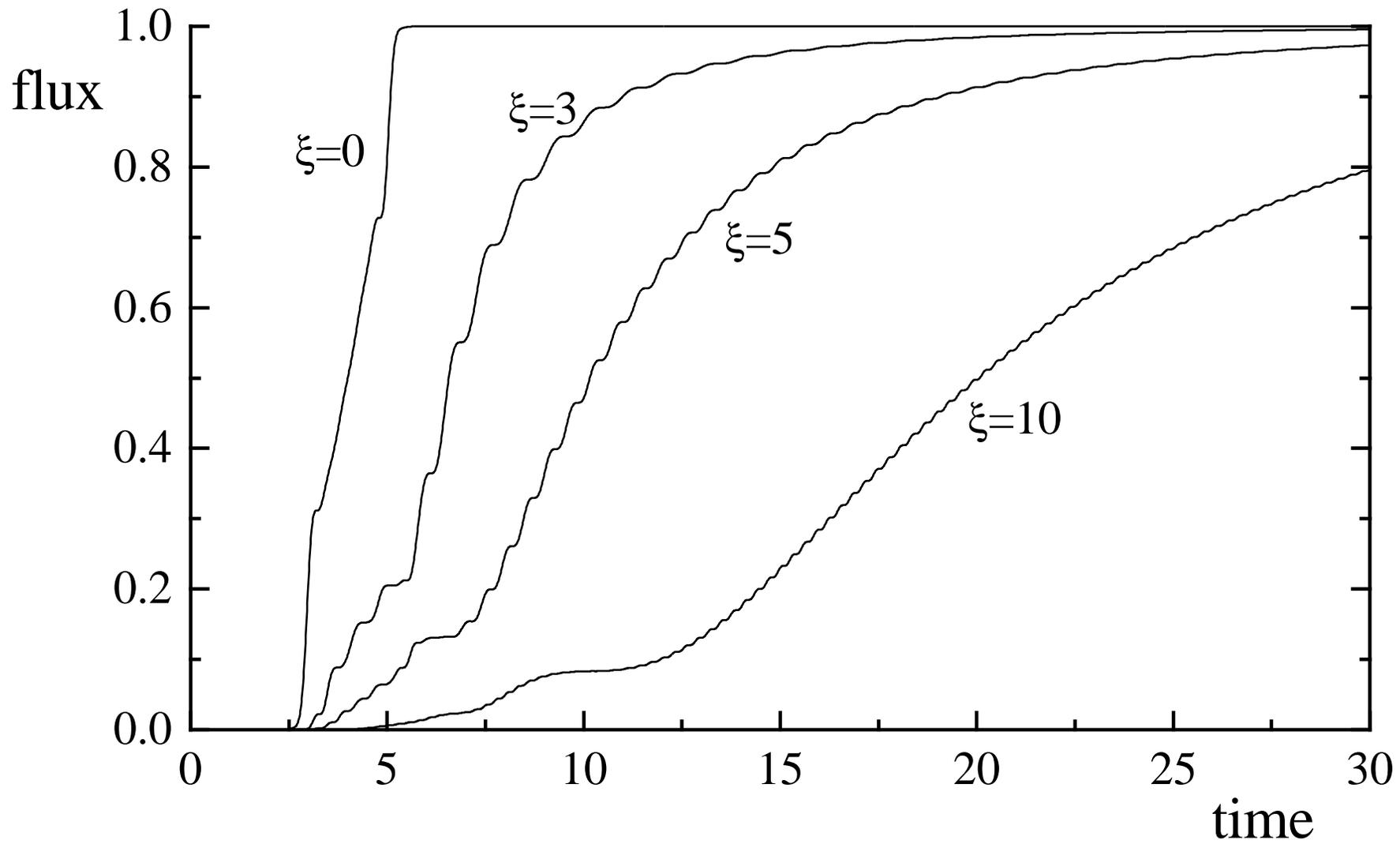
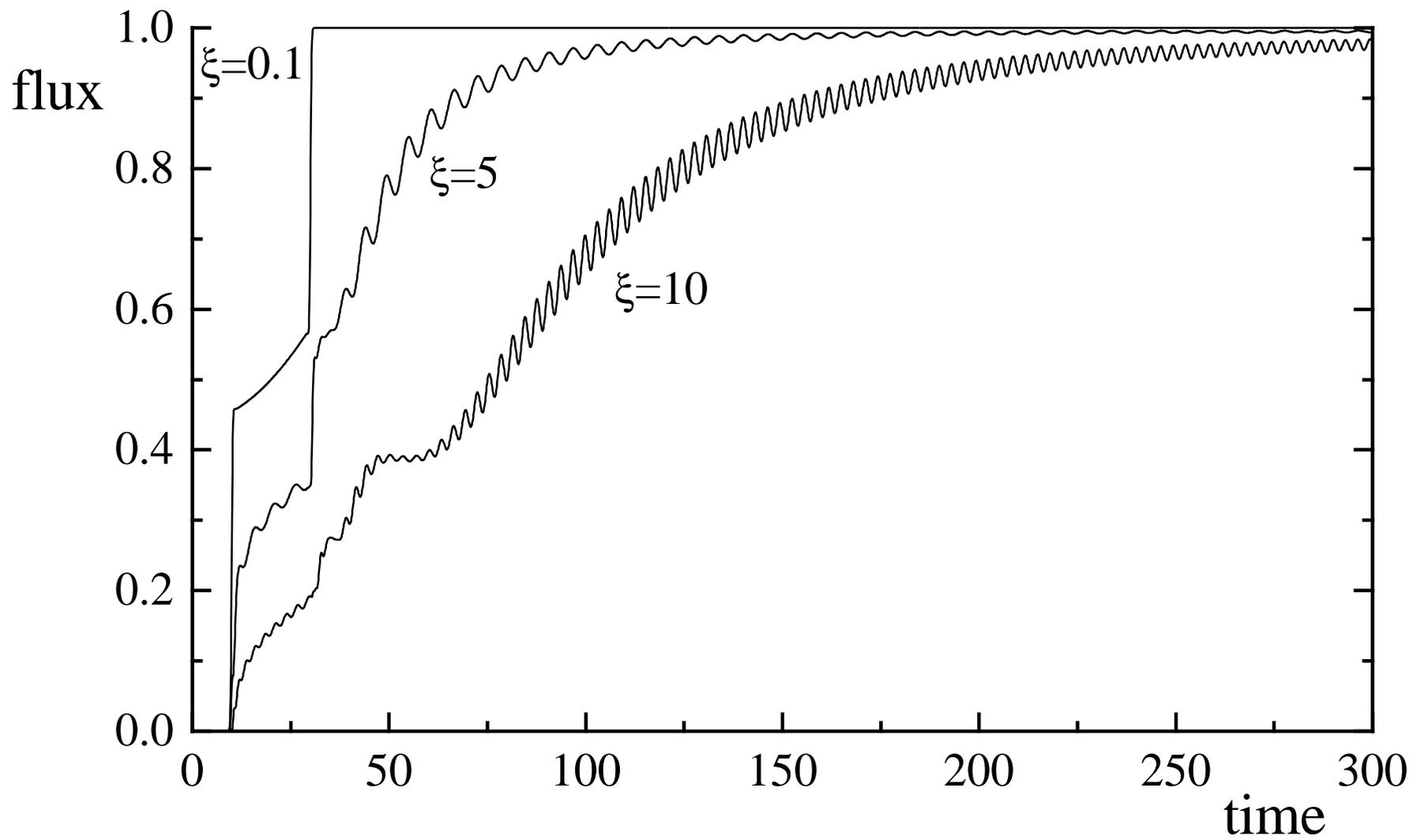
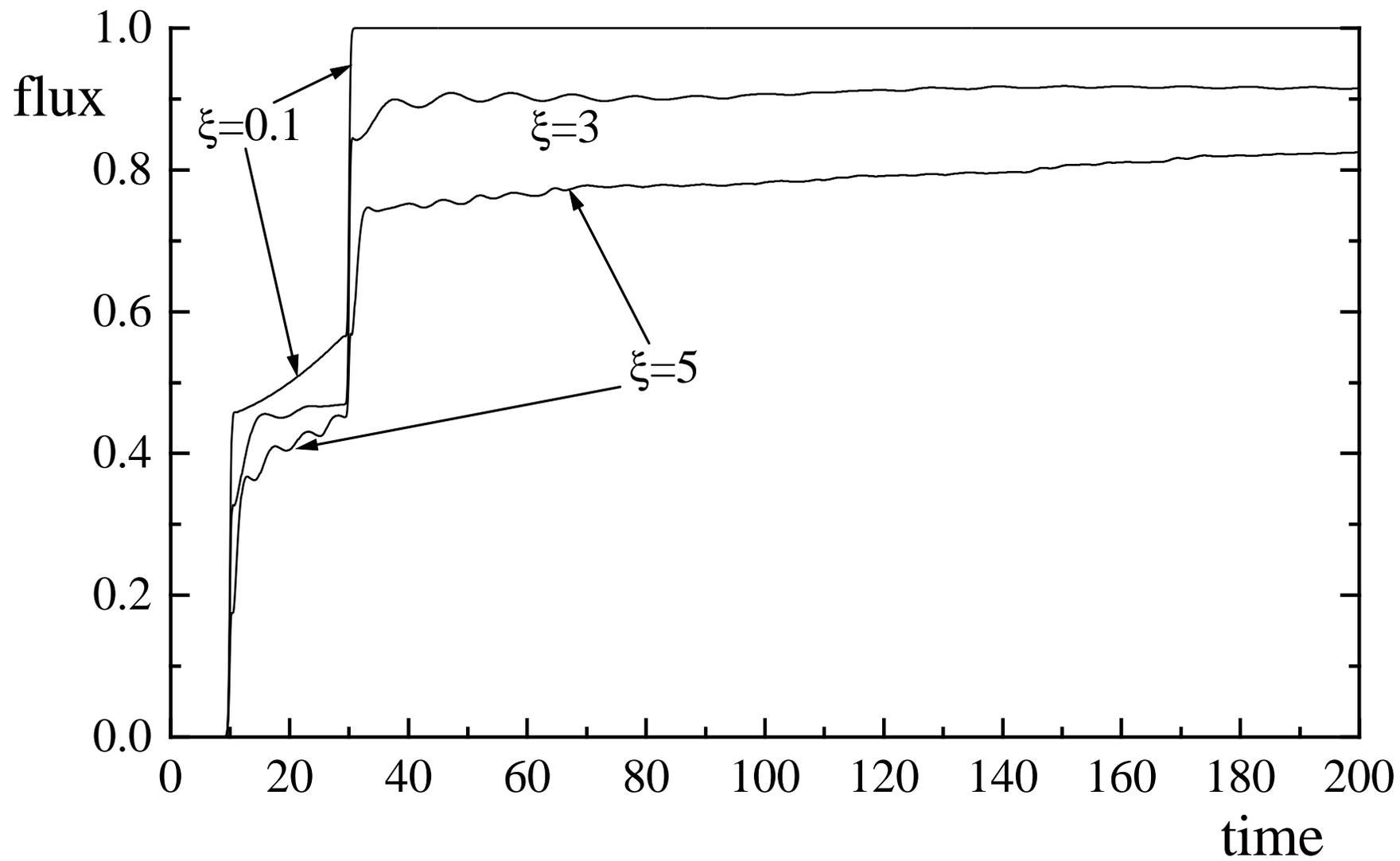
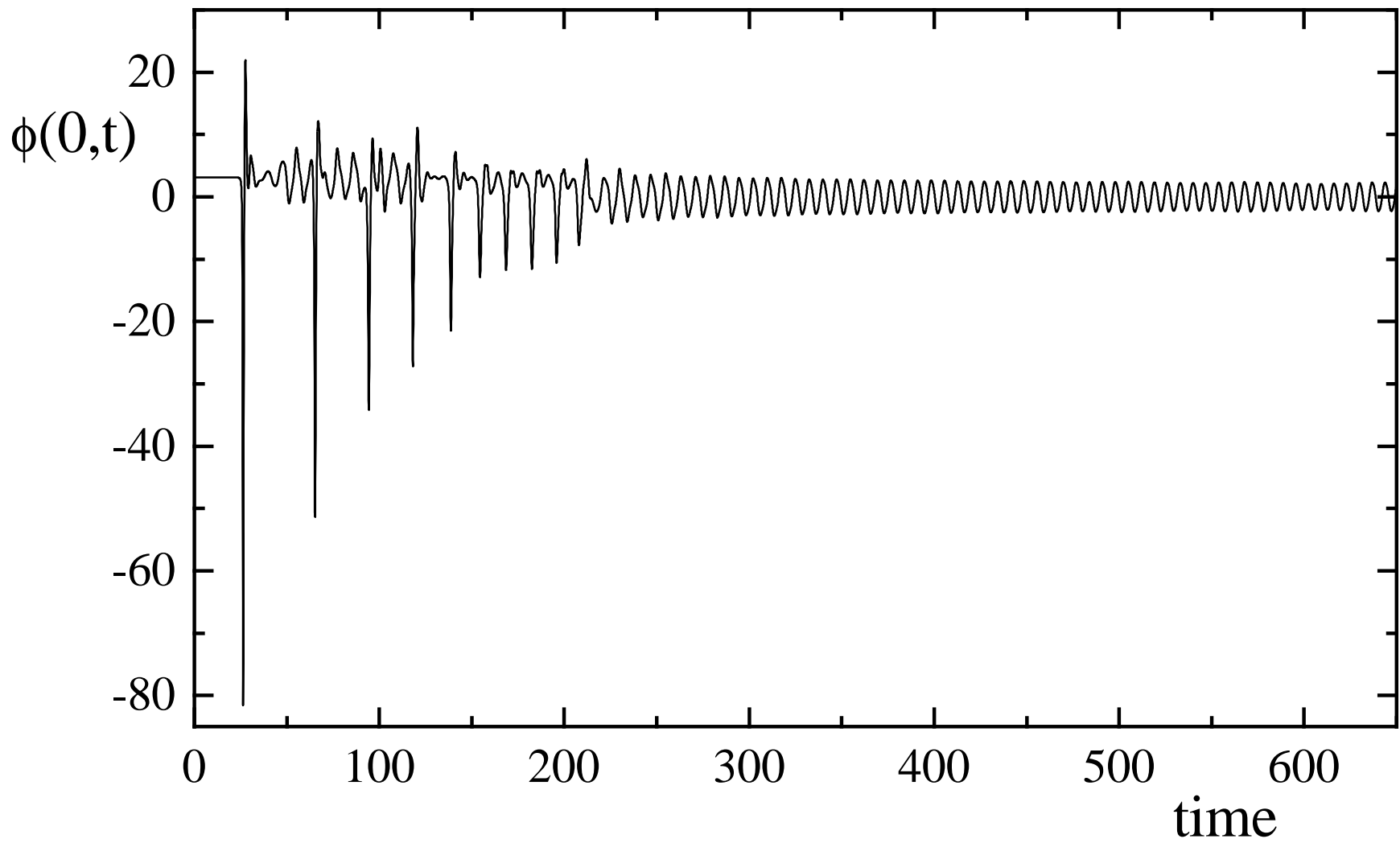
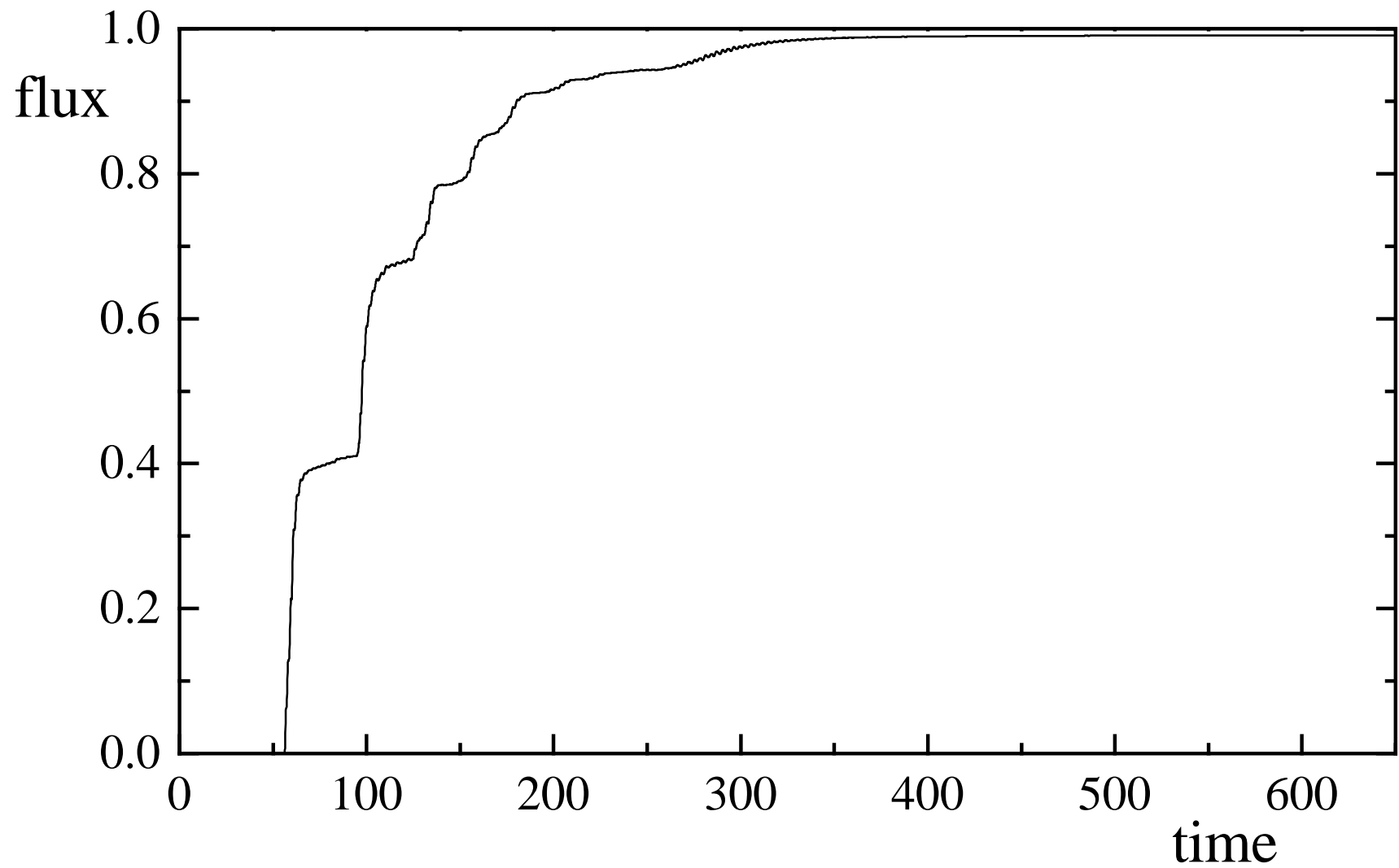
\end{document}